\documentclass[twocolumn,showpacs,preprintnumbers,amsmath,amssymb,superscriptaddress,floatfix,nofootinbib]{revtex4}

\usepackage{graphicx}
\usepackage{amsmath}
\usepackage{amsfonts}
\usepackage{amssymb}

\begin{document}

\title{Hidden charm pentaquark and $\Lambda(1405)$ in the $\Lambda^0_b \to \eta_c K^- p (\pi \Sigma)$ reaction}
\date{\today}
\author{Ju-Jun Xie}~\email{xiejujun@impcas.ac.cn}
\affiliation{Institute of Modern Physics, Chinese Academy of
Sciences, Lanzhou 730000, China} \affiliation{Departamento de
F\'{\i}sica Te\'orica and IFIC, Centro Mixto Universidad de
Valencia-CSIC Institutos de Investigaci\'on de Paterna, Aptdo.
22085, 46071 Valencia, Spain}
\author{Wei-Hong Liang}~\email{liangwh@gxnu.edu.cn}
\affiliation{Department of Physics, Guangxi Normal University,
Guilin 541004, China} \affiliation{Departamento de F\'{\i}sica
Te\'orica and IFIC, Centro Mixto Universidad de Valencia-CSIC
Institutos de Investigaci\'on de Paterna, Aptdo. 22085, 46071
Valencia, Spain}
\author{Eulogio Oset}~\email{oset@ific.uv.es}
\affiliation{Departamento de F\'{\i}sica Te\'orica and IFIC, Centro
Mixto Universidad de Valencia-CSIC Institutos de Investigaci\'on de
Paterna, Aptdo. 22085, 46071 Valencia, Spain} \affiliation{Institute
of Modern Physics, Chinese Academy of Sciences, Lanzhou 730000,
China}

\begin{abstract}

We have performed a study of the $\Lambda^0_b \to \eta_c K^- p$ and
$\Lambda^0_b \to \eta_c \pi \Sigma$ reactions based on the dominant
Cabibbo favored weak decay mechanism. We show that the $K^- p$
produced only couples to $\Lambda^*$ states, not $\Sigma^*$ and that
the $\pi \Sigma$ state is only generated from final state
interaction of $\bar{K}N$ and $\eta \Lambda$ channels which are
produced in a primary stage.  This guarantees that the $\pi \Sigma$
state is generated in isospin $I=0$ and we see that the invariant
mass produces a clean signal for the $\Lambda(1405)$ of higher mass
at $1420$ MeV. We also study the $\eta_c p$ final state interaction,
which is driven by the excitation of a hidden charm resonance
predicted before. We relate the strength of the different invariant
mass distributions and find similar strengths that should be clearly
visible in an ongoing LHCb experiment. In particular we predict that
a clean peak should be seen for a hidden charm resonance that
couples to the $\eta_c p$ channel in the invariant $\eta_c p$ mass
distribution.

\end{abstract}

\maketitle

\section{Introduction}

The analysis of the $\Lambda^0_b \to J/\psi K^- p$ reaction of LHCb
and the interpretation of the $J/\psi p$ spectrum in terms of the
two pentaquark states, $P_c(4380)$ and
$P_c(4450)$~\cite{Aaij:2015tga,Aaij:2015fea} has stirred a wave of
theoretical work trying to understand the nature of the states.
Prior to the experiment there were predictions based on molecular
states of the $\bar{D}\Sigma_c-\bar{D} \Lambda_c$ and $\bar{D}^*
\Sigma_c-\bar{D}^* \Lambda_c$ nature~\cite{Wu:2010jy,Wu:2010vk}.
These would be hidden charm molecular states, and in the case of
$\bar{D}^* \Sigma_c-\bar{D}^* \Lambda_c$ a state with spin-parity
$J^P = 3/2^-$ appears with mass similar to the $P_c(4450)$. These
two systems, studied within coupled channels, couple also to $\eta_c
N$ in the first case and to $J/\psi N$ in the second one, the
channel where the $P_c(4450)$ state was observed. The works of
Refs.~\cite{Wu:2010jy,Wu:2010vk} stimulated further research with
this type of molecular states, studied with different dynamics in
Refs.~\cite{Yang:2011wz,Garcia-Recio:2013gaa,Xiao:2013yca,Uchino:2015uha,Karliner:2015ina},
or quark models~\cite{Yuan:2012wz}, all them prior to the LHCb
experiment. After the experimental observation many ideas have been
proposed to interprete the nature of those pentaquark states.
Molecular states coming from the interaction of meson-baryon have
been
suggested~\cite{Chen:2015loa,Roca:2015dva,He:2015cea,Huang:2015uda,Meissner:2015mza,
Xiao:2015fia,Eides:2015dtr,Yang:2015bmv,Chen:2016heh,Lu:2016nnt,Shimizu:2016rrd,Shen:2016tzq,
Xiao:2016ogq,Chen:2016ryt,Wu:2017weo,Shimizu:2017xrg,Yamaguchi:2017zmn}.
Pentaquark structures of type diquark-diquark-antiquark nature have
also been proposed~\cite{Lebed:2015tna,Zhu:2015bba}. Other different
quark rearrangements have been
suggested~\cite{Mironov:2015ica,Gerasyuta:2015djk,Santopinto:2016pkp,Ortega:2016syt},
as well as QCD sum rules~\cite{Chen:2015moa,Azizi:2016dhy}, and new
methods of production in different reactions have also been
investigated~\cite{Garzon:2015zva,Blin:2016dlf,Ali:2016dkf,Kubarovsky:2016whd,Liu:2016dli,Chen:2015sxa,Feijoo:2015kts,Lu:2016roh,Lin:2017mtz}.
The properties of these hidden charm states in light quark matter
have also been studied in Ref.~\cite{Cleven:2017fun}. Reviews on the
subject have also been written with detailed discussion of different
works and ideas in
Refs.~\cite{Burns:2015dwa,Chen:2016qju,Esposito:2016noz,Oset:2016nvf,Ali:2017jda,Zhou:2017bhq,Lu:2017dvm,Olsen:2017bmm,Guo:2017jvc,Lebed:2016hpi}.

A suggestion to explain the $P_c(4450)$ peak as a manifestation of a
triangle singularity~\cite{Guo:2015umn,Liu:2015fea} was shown in
Ref.~\cite{Bayar:2016ftu} to be unable to explain the experimental
feature with the preferred quantum numbers of the experiment $3/2^-$
or $5/2^+$.

Prior also to the experimental measurement of the $\Lambda^0_b \to
J/\psi K^- p$ reaction~\cite{Aaij:2015fea,Aaij:2015tga}, a
theoretical study was done in Ref.~\cite{Roca:2015tea}, where it was
shown that the $K^- p$ was produced in isospin $I =0$ (as was later
on corroborated by experiment) and also the invariant $K^- p$ mass
distribution in $s$-wave (related to the $\Lambda(1405)$ production)
and the invariant $\pi \Sigma$ mass distribution in the related
$\Lambda^0_b \to J/\psi \pi \Sigma$ reaction, were studied.

The experimental analysis of the $\Lambda^0_b \to J/\psi K^- p$
reaction~\cite{Aaij:2015fea,Aaij:2015tga} also showed the
contribution of the $K^- p$ mass distribution from the
$\Lambda(1405)$, which was in qualitative agreement with the one
found in Ref.~\cite{Roca:2015tea}. After the experiment was done,
the consistency of the strength of the peak of the $P_c(4450)$ and
the $K^-p$ strength coming from the $\Lambda(1405)$, were shown to
be consistent with the findings of Refs.~\cite{Wu:2010jy,Wu:2010vk}
should the quantum number be $1/2^-$~\cite{Roca:2015dva}, but this
was generalized to other quantum numbers in
Ref.~\cite{Roca:2016tdh}. Incidentally, in this latter work it was
shown that, based on the $J/\psi p$ and $K^- p$ mass distributions
alone, one could not determine the spin and parity of the states,
nor the need for the wide $P_c(4380)$ state, which means that
angular distributions and polarizations information must be the
elements helping determining these quantum numbers in the
experimental analysis.

In the present work we pay attention to the reaction of $\Lambda^0_b
\to \eta_c K^- p$, which is under analysis by the LHCb
collaboration~\cite{liming} and make predictions for the $\eta_c p$
and $K^- p$ mass distributions. Simultaneously, we also study the
$\Lambda^0_b \to \eta_c \pi \Sigma (\pi \Sigma \equiv \pi^+
\Sigma^-, \pi^0 \Sigma^0, \pi^- \Sigma^+)$ reaction and make
predictions for the $\pi \Sigma$ mass distribution which shows the
$\Lambda(1405)$ shape. We use the predictions made for the
$\bar{D}\Sigma_c-\bar{D}\Lambda_c$ and coupled channels in
Refs.~\cite{Wu:2010jy,Wu:2010vk} and can relate the $\pi \Sigma$
mass distribution with those of $\eta_c p$ and $K^- p$. The
interesting thing is that a clear peak emerges in the $\eta_c p$
mass distribution due to a $1/2^-$ dynamically generated state,
mostly for $\bar{D} \Sigma_c$, which couples relatively strongly to
the $\eta_c p$ channel. The predictions done here should be of much
use to guide experimental search and to get relevant conclusion from
a comparison with experiment when this is finished.

This article is organized as follows. In Sec.~\ref{Sec:Formalism},
we present the theoretical formalism of the decay of $\Lambda^0_b
\to \eta_c K^- p$, explaining in detail the hadronization and final
state interactions of the $\eta_c p$ and $K^-p$ pairs. Numerical
results and discussions are presented in Sec.~\ref{Sec:Results},
followed by a summary in the last section.

\section{Formalism} \label{Sec:Formalism}

\begin{figure}[htbp]
\begin{center}
\includegraphics[scale=0.9]{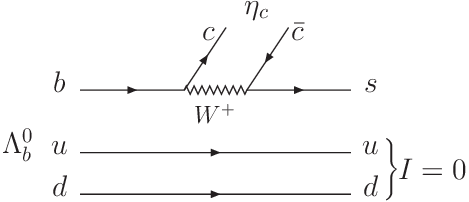}
\caption{Diagrammatic representation of the $\Lambda^0_b \to \eta_c
K^- p$ decay at the quark level prior the hadronization of the final
state. \label{fig:fig1}}
\end{center}
\end{figure}

Following Ref.~\cite{Roca:2015tea} we write the first step in the
$\Lambda^0_b \to \eta_c K^- p$ reaction at the quark level, which
proceeds as shown in Fig.~\ref{fig:fig1}. The $ud$ diquark in the
$\Lambda^0_b$ is in $I = 0$ and they are spectators in the decay.
The final state $sud$ is again in $I=0$ and hence, only $\Lambda^*$
states should show up in the final state apart of the $c \bar{c}$
that now forms the $\eta_c$. Note that, apart from the $bcW$
coupling in the first weak vertex, the next coupling $csW$ is
Cabibbo favored.

\begin{figure}[htbp]
\begin{center}
\includegraphics[scale=0.9]{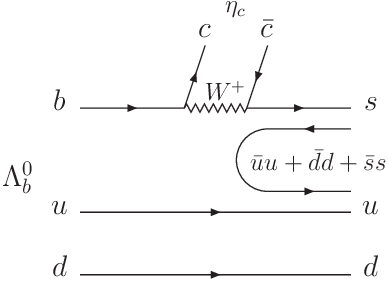}
\caption{Hadronization of the final state $sud$ of
Fig.~\ref{fig:fig1} including the production of $\bar{u} u + \bar{d}
d + \bar{s}s$. \label{fig:fig2}}
\end{center}
\end{figure}

We must now proceed to produce $K^- p$ from the $sud$ cluster of the
final state and we are interested in $K^- p$ in $s$-wave, which is
what couples to the $\Lambda(1405)$, the dominant term close to the
$K^-p$ threshold, as shown in the experimental analysis of the
$\Lambda^0_b \to J/\psi K^- p$
reaction~\cite{Aaij:2015fea,Aaij:2015tga}. Since $K^- p$ in $s$-wave
has negative parity and the $u$, $d$ quarks are spectators, the $s$
quark must be produced in orbital angular momentum $L=1$ in the
diagram of Fig.~\ref{fig:fig1}. Yet, since finally in $K^- p$ all
quarks are in the ground state, the hadronization, introducing a
$q\bar{q}$ pair with the quantum number of the vacuum, must involve
the $s$ quark to bring it back to the ground state. This is shown in
Fig.~\ref{fig:fig2}.

The details of the hadronization are shown in
Ref.~\cite{Roca:2015tea}, with the resulting hadronic structure
$|H>$ given by~\footnote{Note that the $|\eta \Lambda>$ and $|\eta'
\Lambda>$ terms have a different sign than in
Ref.~\cite{Roca:2015tea}. This is due to the fact that in
Ref.~\cite{Roca:2015tea} the prescription of Close~\cite{close} for
the baryon states in terms of quarks was used. However, when using
chiral Lagrangians, as we do here, one has to adhere to a different
sign convention which is shown in Table III of the
work~\cite{Miyahara:2016yyh}. The $\Lambda$, $\Sigma^+$, and $\Xi^0$
states have opposite sign to those of
Close~\cite{Pavao:2017cpt,Miyahara:2016yyh}.}
\begin{equation}\label{eq:mesonbaryon}
|H> \equiv |K^-p> + |\bar{K}^0 n> + \frac{\sqrt{2}}{3}|\eta \Lambda>
- \frac{2}{3}|\eta' \Lambda> .
\end{equation}

As in Ref.~\cite{Roca:2015tea} we neglect the $\eta' \Lambda$
channel in our study since it has a much larger mass than $\eta
\Lambda$ or $K^- p$.

\begin{figure}[htbp]\centering
\includegraphics[scale=0.65]{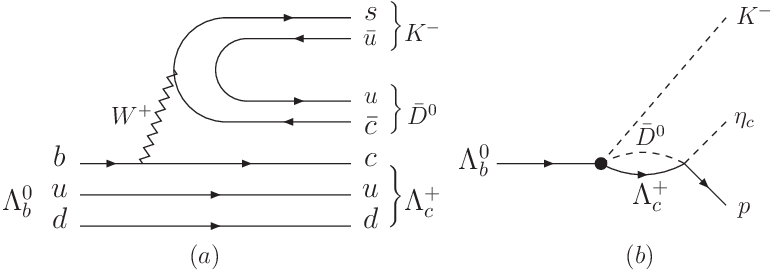}
\caption{A different mechanism for the $\Lambda^0_b \to \eta_c K^-
p$ reaction. \label{fig:fig3}}
\end{figure}

We should note that the $\eta_c p K^-$ final state can be produced
in a different way as shown in Fig.~\ref{fig:fig3}. The mechanism
proceeds via $\bar{c} s$ production via external
emission~\cite{Chau:1982da} followed by hadronization via $\bar{u}u$
creation, producing $K^- \bar{D}^0 \Lambda^+_c$ as shown in
Fig.~\ref{fig:fig3} (a). The $\bar{D}^0 \Lambda^+_c$ undergo final
state interaction to produce $\eta_c p$ as shown in
Fig.~\ref{fig:fig3} (b). Yet, this mechanism is much suppressed due
to the fact that it involves the product of the $\bar{D}^0
\Lambda^+_c$ and $\eta_c p$ couplings to the resonance, $R$, that is
found in Ref.~\cite{Wu:2010jy,Wu:2010vk}. Indeed, these couplings
are $g_{R,\bar{D}\Lambda_c} = -0.08 + {\rm i}0.06$ and $g_{R,\eta_c
p} = -0.94 + {\rm i}0.03$ compared to the $\bar{D}\Sigma_c$ coupling
of $g_{R,\bar{D}\Sigma_c} = 2.96 - {\rm i}0.21$. The mechanism that
we study here to produce the hidden charm resonance, through
rescattering of the $\eta_c p$ state has much larger strength than
the mechanism of Fig.~\ref{fig:fig3} and we disregard this latter
one.

\begin{figure*}[htbp]\centering
\includegraphics[scale=0.85]{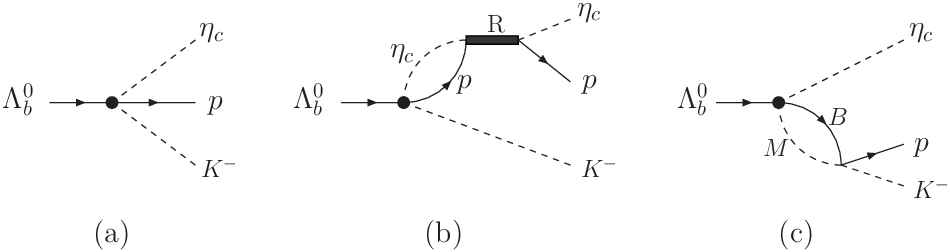}
\caption{Diagrammatic representation of the final state interaction
of the meson-baryon components of $|H>$ in
Eq.~\eqref{eq:mesonbaryon}: (a) direct $\eta_c K^- p$ vertex at tree
level, (b) final state interaction of $\eta_c p$, and (c) final
state interaction of $K^- p$. $MB$ stands for $K^-p$, $\bar{K}^0 n$,
and $\eta \Lambda$. \label{fig:fig4}}
\end{figure*}

The last step to generate the $\eta_c p K^-$ involves final state
interaction of the meson-baryon components of $|H>$ in
Eq.~\eqref{eq:mesonbaryon}. This is depicted diagrammatically in
Fig.~\ref{fig:fig4}, where $\eta_c p$ and $K^- p$ final state
interactions are considered.

In Fig.~\ref{fig:fig4} we consider the final state interaction of
$\eta_c p$ because there is a resonance $R$ generated by
$\bar{D}\Sigma_c$, $\bar{D}\Lambda_c$, and $\eta_c N$ in
Refs.~\cite{Wu:2010jy,Wu:2010vk} and the $\eta_c N$ is one of the
channels that has a relatively large coupling to this resonance. In
Ref.~\cite{Wu:2010jy} we find a state of $I = 1/2$, $J^P = 1/2^-$,
with
\begin{eqnarray}
M_R - i\frac{\Gamma_R}{2} = (4265 - i 11.6) ~ {\rm MeV}.
\end{eqnarray}

Then the transition matrix for $\Lambda^0_b \to \eta_c p K^-$ in
Fig.~\ref{fig:fig4} is given by,
\begin{eqnarray}
t &=& V_P \biggl ( 1 + G_{\eta_c p}(M_{\eta_c p}) t_{\eta_c p \to
\eta_c
p}(M_{\eta_c p}) \nonumber  \\
&& + \sum_i h_i G_i(M_{K^- p}) t_{i \to K^- p}(M_{K^- p}) \biggr ),
\label{eq:totalamplitude}
\end{eqnarray}
where $h_i$ is the weight of the production of the different
meson-baryon states in Eq.~\eqref{eq:mesonbaryon},
\begin{eqnarray}
h_{K^- p} = 1, ~~~ h_{\bar{K}^0 n} = 1, ~~~ h_{\eta \Lambda} =
\frac{\sqrt{2}}{3}.
\end{eqnarray}

In Eq.~\eqref{eq:totalamplitude}, $G_{\eta_c} p$ is the $\eta_c p$
loop function, which depends on the invariant mass $M_{\eta_c}p$ of
the final $\eta_c p$ system, while $G_i$ ($i = K^- p$, $\bar{K}^0
n$, and $\eta \Lambda$) denotes the meson-baryon loop function,
which depends on the invariant mass $M_{K^- p}$ of the final $K^- p$
system. The factor $V_P$ is the strength of the tree level
$\Lambda^0_b \to \eta_c K^- p$, which is unknown in our approach.
This means we will only look at invariant mass distributions
relative to each other.

The amplitude $t_{\eta_c p \to \eta_c p}$ is given by
\begin{eqnarray}
t_{\eta_c p \to \eta_c p}(M_{\eta_c p}) = \frac{g^2_{R,\eta_c
p}}{M^2_{\eta_c p} - M^2_R + i M_R \Gamma_R}, \label{eq:tetacp}
\end{eqnarray}
and $t_{i \to K^- p}$ ($i = K^- p$, $\bar{K}^0 n$, and $\eta
\Lambda$) are the transition matrix elements evaluated with the
chiral unitary approach in Ref.~\cite{Oset:1997it}. The $t$ matrix
is given in terms of the Bethe-Salpeter equation by
\begin{eqnarray}
t = [1-VG]^{-1} V,
\end{eqnarray}
with $V$ the transition potential evaluated from the chiral
Lagrangians~\cite{Ecker:1994gg} and $G$ the loop function for the
intermediate meson-baryon states, which is the same appearing in
Eq.~\eqref{eq:totalamplitude}. We use the same as in
Ref.~\cite{Oset:1997it} with cut off regularization and a cut off,
$q_{\rm max} = 630$ MeV. As for the $G_{\eta_c p}$ loop function we
use the same as in Ref.~\cite{Wu:2010jy}, which in this case is done
using dimensional regularization with the scale parameter $\mu =
1000$ MeV and the subtraction constant $a_{\mu} = -2.3$.

It is obvious that with the phase space available for $K^- p$
production one obtains a large range of invariant masses that
accommodates the excitation of many $\Lambda^*$ states, as in the
$\Lambda^0_b \to J/\psi K^- p$ reaction of
Refs.~\cite{Aaij:2015fea,Aaij:2015tga} (see also the alternative
analysis in Ref.~\cite{Roca:2016tdh}). This means that in the $K^-
p$ invariant mass distribution we aim at getting only the mass
distribution close to $K^- p$ threshold in $s$-wave. The $\eta_c p$
interaction is OZI suppressed  and it is only relevant close to the
pole of $R$. Yet, since the $\Lambda^*$ excitation reverts into the
$\eta_c p$ mass distribution, we will also pay not much attention to
the strength of the background in $\eta_c p$ but to the strength of
the peak.

The double mass differential width when one sums and averages the
polarizations of the particles is given by~\cite{Patrignani:2016xqp}
\begin{eqnarray}
\frac{d^2\Gamma}{dM_{\eta_c p} dM_{K^- p}}= \frac{m_p M_{\eta_c
p}M_{K^- p}}{16\pi^3 M^2_{\Lambda^0_b}} |t|^2. \label{eq:dgdmdm}
\end{eqnarray}
By integrating over $M_{K^- p}$ in Eq.~\eqref{eq:dgdmdm} we obtain
$d\Gamma/dM_{\eta_c p}$. The limits of integration are found in
Ref.~\cite{Patrignani:2016xqp}. Similarly, we can obtain the limits
of $M_{\eta_c p}$ when we fix $M_{K^- p}$. By integrating over
$M_{\eta_c p}$ in Eq.~\eqref{eq:dgdmdm} we obtain $d\Gamma/dM_{K^-
p}$. In this way Eq.~\eqref{eq:dgdmdm} provides a Dalitz plot and
$d\Gamma/dM_{\eta_c p}$, $d\Gamma/dM_{K^- p}$ the projection over
the $\eta_c p$ and $K^- p$ invariant masses.

For the $\Lambda^0_b \to \eta_c \pi \Sigma$ reaction, unlike the
$\Lambda^0_b \to \eta_c p K^-$ which can be produced at tree level
[see Fig.~\ref{fig:fig4} (a)] without final state interactions (see
$|H>$ in Eq.~\eqref{eq:mesonbaryon}, the $\eta_c \pi \Sigma$ states
does not appear at tree level since it is not contained in $|H>$).
The only way to get it is through final state interaction of the
meson-baryon components of $|H>$ in Eq.~\eqref{eq:mesonbaryon}. This
can be done with the mechanism shown in Fig.~\ref{fig:fig4} (c) by
replacing $K^- p$ with $\pi \Sigma$.

Then we find
\begin{eqnarray}
t' &=& V_P \sum_i G_i (M_{\pi \Sigma}) t_{i \to \pi \Sigma} (M_{\pi
\Sigma}). \label{eq:tpisigma}
\end{eqnarray}
We can use the same formulas as before changing $K^- p$ by $\pi
\Sigma$ in the final state, and $V_P$ is the same as in the former
reaction,~\footnote{In this work, we take $V_P = 1 {\rm MeV}^{-1}$.}
which allows us to compare the different mass distributions. The
amplitudes $t_{i \to \pi \Sigma}$, as well as $G_i$ are calculated
with the chiral unitary approach of Ref.~\cite{Oset:1997it} as
before.

\section{Numerical results} \label{Sec:Results}

\begin{figure}[htbp]\centering
\includegraphics[scale=0.45]{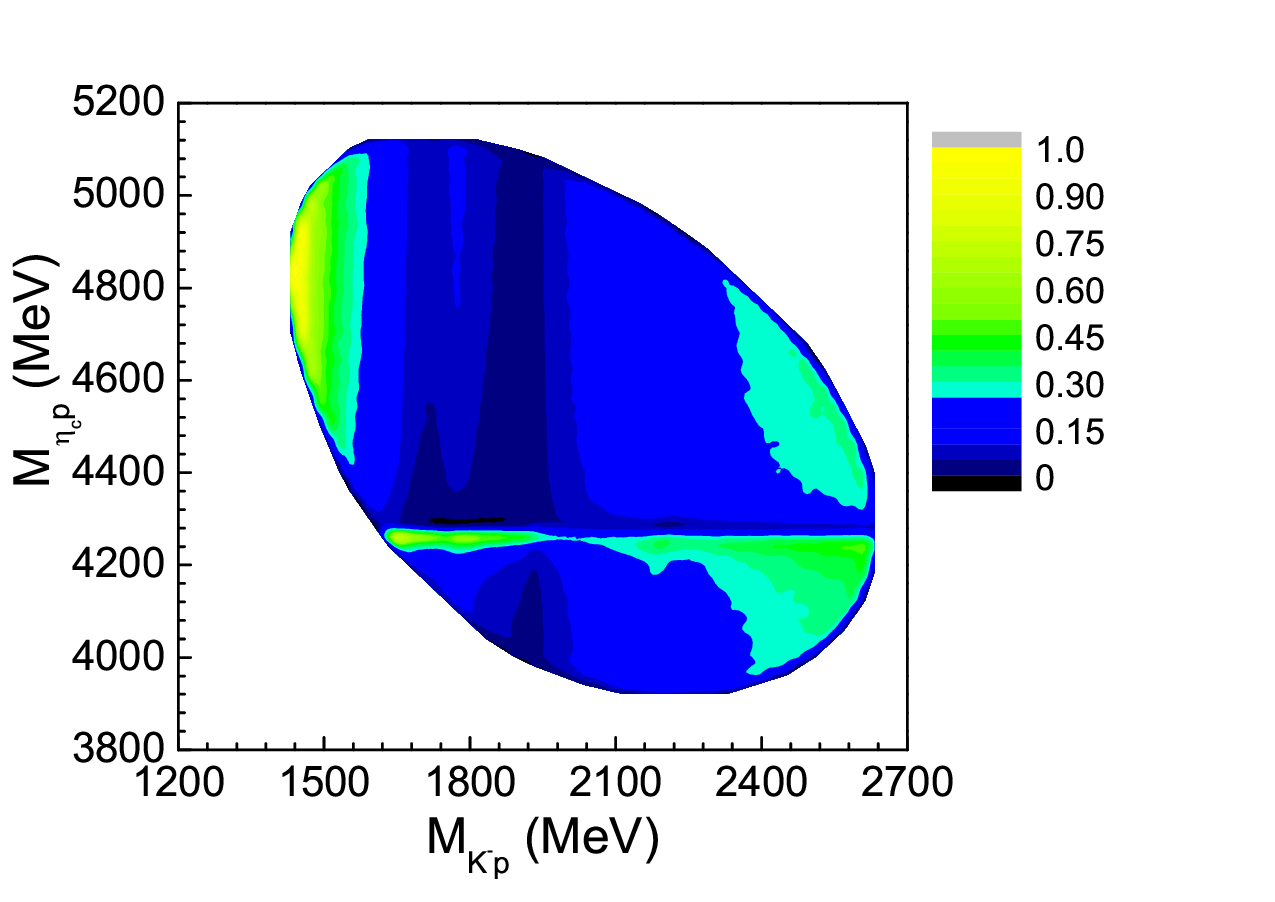}
\caption{(Color online) Dalitz plot representation for the invariant
masses of $\eta_c p$ and $K^- p$. \label{fig:dp}}
\end{figure}

In Fig.~\ref{fig:dp}, we show the Dalitz plot for the invariant
masses of $\eta_c p$ and $K^- p$. In the figure we can see clearly
the signals for the $\Lambda(1405)$ in $K^- p$, close to the $K^- p$
threshold and in $\eta_c p$ for the resonance $R$.

\begin{figure}[htbp]\centering
\includegraphics[scale=0.42]{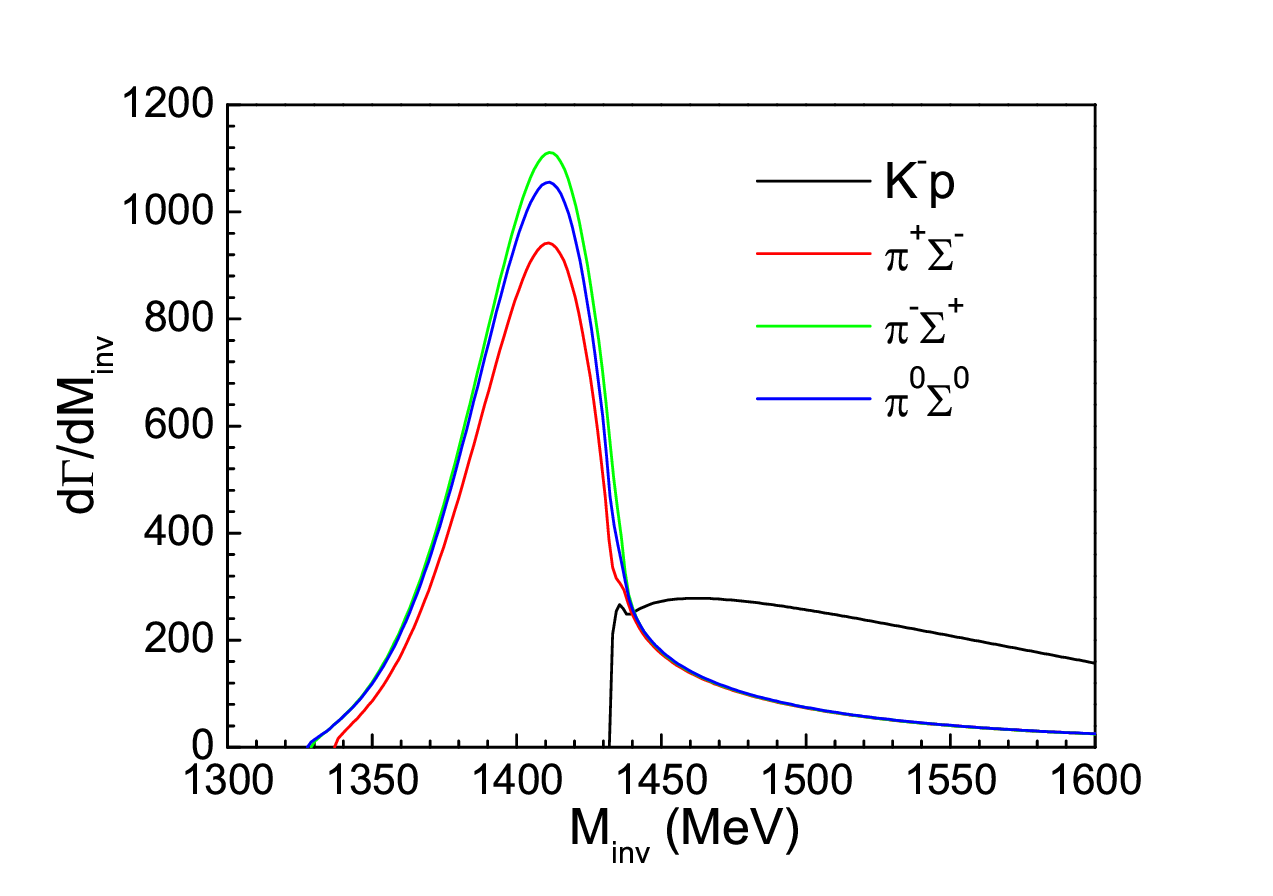}
\caption{(Color online) The $K^- p$ and $\pi \Sigma$ invariant
masses distributions. \label{fig:dgdm-MB}}
\end{figure}

In Fig.~\ref{fig:dgdm-MB} we show the invariant mass distribution
for $K^- p$ ($M_{\rm inv} \equiv M_{K^- p}$) and $\pi \Sigma$
($M_{\rm inv} \equiv M_{\pi \Sigma}$). We stress once more that the
$K^- p$ is only for $s$-wave, related to the $\Lambda(1405)$
production close to threshold. We can expect extra strength from
$\Lambda(1520)$ excitation and other resonances, but with the
partial wave anlysis of LHCb one can separate the contributions of
different resonances as done for the $\Lambda^0_b \to J/\psi K^- p$
reaction~\cite{Aaij:2015fea,Aaij:2015tga}, and compare with our
results. Very interesting is to compare the strength and shape of
$\pi \Sigma$ production with $K^- p$. The results for the $\pi
\Sigma$ mass distribution deserve some attention. The three
distributions for $\pi^+ \Sigma^-$, $\pi^0 \Sigma^0$, and $\pi^-
\Sigma^+$ are very similar, they peak at the same energy and appear
with no background. This is a consequence of the dynamics of their
production. Indeed, in Eq.~\eqref{eq:mesonbaryon} we see that $\pi
\Sigma$ is not produced at tree level. It is only produced by
rescattering as seen in Eq.~\eqref{eq:tpisigma}. This means that the
$\Lambda(1405)$ resonance is produced clearly without background
from tree level. Second, since we saw that in the final meson-baryon
states we had $I =0$, this means that the $\pi \Sigma$ is produced
in $I =0$ without contribution of $I=1$, for instance the
$\Sigma(1385)$ or other $I =1$ background sources. This is actually
a problem in many reactions producing the $\Lambda(1405)$, as
photoproduction~\cite{Moriya:2013hwg} or the $pp \to pK^- \pi
\Sigma$ reaction~\cite{Agakishiev:2012xk}. One good consequence of
this is that the different $\pi^+ \Sigma^-$, $\pi^0 \Sigma^0$, and
$\pi^- \Sigma^+$ are produced with similar strength and peaking at
the same place, which does not occur when there is contribution from
both $I =0$ and $I=1$~\cite{Nacher:1998mi}. There is another
interesting feature which is that all these distributions peak at
$1420$ MeV. This is a consequence of the dynamics of the
$\Lambda(1405)$ and the two states (two poles in the same Rieman
sheet) that are associated to it~\cite{Oller:2000fj,Jido:2003cb}. In
these chiral pictures, corroborated by all works in chiral
dynamics,~\footnote{See the chapter "{\it pole structure of the
$\Lambda(1405)$ region}" of the PDG~\cite{Patrignani:2016xqp}.}
there are two states, one with mass around $1420$ MeV, that couples
strongly to $\bar{K}N$, and the other one at around $1385$ MeV that
couples mostly to $\pi \Sigma$. The dynamics of the present reaction
is such that the $\Lambda(1405)$ is initially produced by $\bar{K}
N$ (see Eq.~\eqref{eq:totalamplitude} and Eq.~\eqref{eq:tpisigma}),
hence, it is basically the $\Lambda^*$ state at $1420$ MeV the one
which is excited, and this is seen in Fig.~\ref{fig:dgdm-MB}. This
selection of the upper $\Lambda^*$ states also occur, and is
supported experimentally, in other reactions where the
$\Lambda(1405)$ is initiated by the $\bar{K} N$ channel, as the $K^-
p \to \pi^0 \pi^0 \Sigma^0$
reaction~\cite{Prakhov:2004an,Magas:2005vu} or the $K^- d \to n \pi
\Sigma$ reactions~\cite{Braun:1977wd,Jido:2009jf}. The discussion
done here indicates that the present reaction, $\Lambda^0_b \to
\eta_c \pi \Sigma$, is an ideal one to show in a very clean way the
upper state of the two $\Lambda(1405)$ states.

\begin{figure}[htbp]\centering
\includegraphics[scale=0.42]{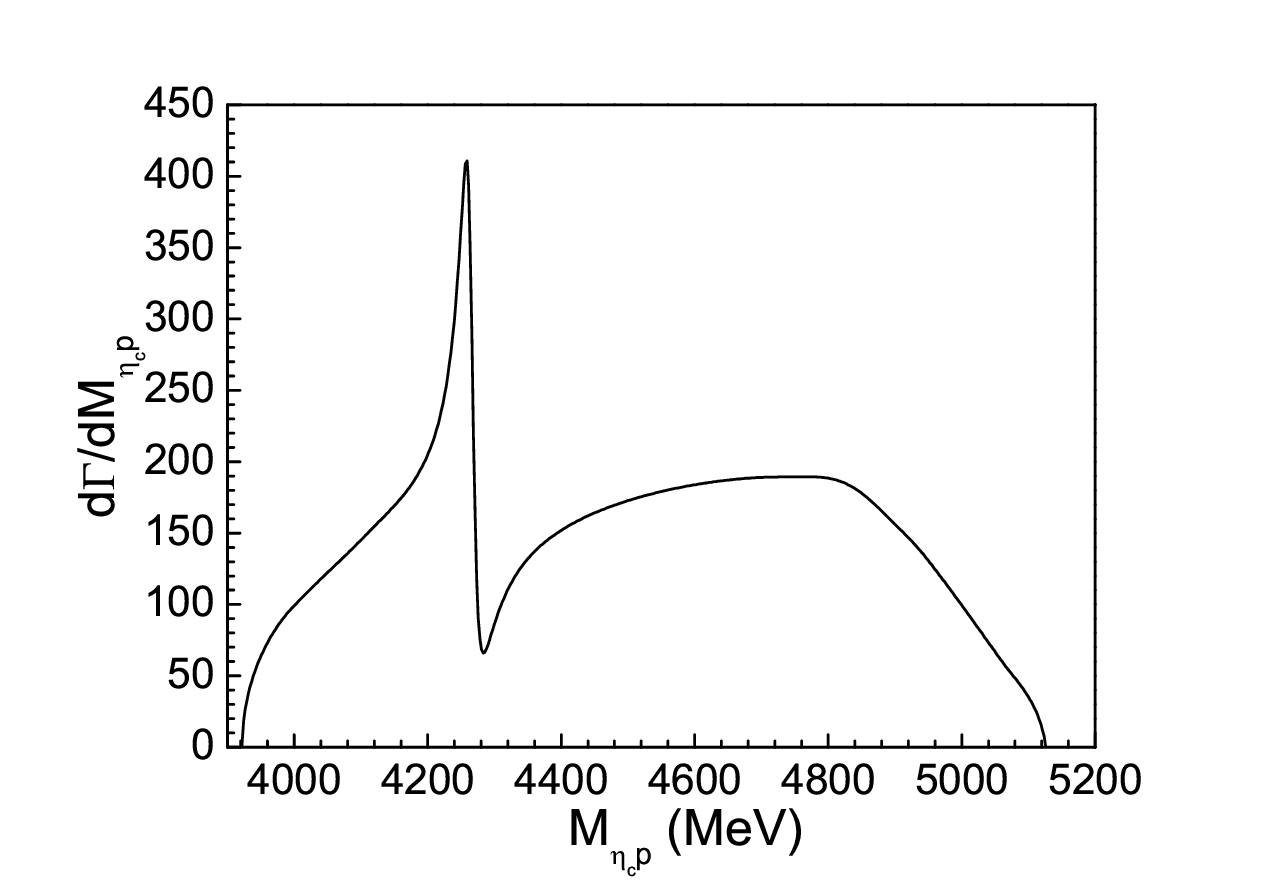}
\caption{The $\eta_c p$ invariant mass distribution.
\label{fig:dgdm-etacp}}
\end{figure}

Finally, in Fig.~\ref{fig:dgdm-etacp}, we show the mass distribution
of $\eta_c p$. We see a strong and clear peak around the mass of the
dynamically generated hidden charm resonance $M_R = 4265$ MeV. The
peak has a large strength, bigger than the $K^- p$ strength at the
peak, which indicates that it should be clearly visible. This is the
comparison we want to make, and not the comparison of the strengths
of the peak with the background, because the background in
Fig.~\ref{fig:dgdm-etacp} is obviously underestimated since we do
not consider the excitation of other $\Lambda^*$ apart from the
$\Lambda(1405)$, which would fill the region below the peak in
Fig.~\ref{fig:dgdm-etacp} with extra background. We should also warn
that the mass of $R$ in Refs.~\cite{Wu:2010jy,Wu:2010vk} is a
prediction, but one has uncertainties in the mass, tied to the
choice of the subtraction constant. Uncertainties of about 20 MeV,
or even more, are expected, but the stability of the strength of the
peak has been studied in similar reactions producing hidden charm
states~\cite{Chen:2015sxa,Feijoo:2015kts,Lu:2016roh} and the same
should happen here.

We have not cared about the absolute normalization in the work.
However, there is an interesting exercise that we can do. Since
$\eta_c$ and $J/\psi$ are both $c\bar{c}$ states which differ only
in the spin alignments, we can use heavy quark spin symmetry (HQSS)
to relate the reactions $\Lambda^0_b \to J/\psi K^- p$ and
$\Lambda^0_b \to \eta_c K^- p$. Semileptonic decays have been
investigated within the HQSS
formalism~\cite{Jenkins:1992nb,Lu:1995ug}. The nonleptonic decay
with the internal emission topology is more complicated, because it
has two quark vertices, rather than one in the semileptonic decay.
We have done our own formulation of the problem, using Racah algebra
and we show the derivation in the Appendix. The conclusion is that
the rate of $\Lambda^0_b \to \eta_c K^- p$ production with $K^- p$
in $S$-wave is three times bigger than for $\Lambda^0_b \to J/\psi
K^- p$ apart from phase space. This information is useful because
the $\Lambda^0_b \to J/\psi K^- p$ has been investigated in
LHCb~\cite{Aaij:2015fea,Aaij:2015tga} and thus we should expect
strengths for the $\Lambda^0_b \to \eta_c K^- p$ reaction reasonably
bigger than for $\Lambda^0_b \to J/\psi K^- p$.

\section{Summary and conclusions}

We have performed a study of the $\Lambda^0_b \to \eta_c K^- p$ and
$\Lambda^0_b \to \eta_c \pi \Sigma (\pi^+ \Sigma^-, \pi^0 \Sigma^0,
\pi^- \Sigma^+)$ reactions. We identify the mechanism for the
reaction at quark level and see that the $K^- p$ produced couples
only to $\Lambda^*$ states and not $\Sigma^*$ states. The Cabibbo
favored mechanism (up to the $bcW$ vertex, necessary for the weak
decay) produces an $sud$ cluster in the final state that, upon
hadronization, leads to $K^- p$, $\pi^+ \Sigma^-$, $\pi^0 \Sigma^0$,
and $\pi^- \Sigma^+$ in the final state, and this interaction is
basically mediated by the $\Lambda(1405)$ state of high mass at
$1420$ MeV, such that the different $\pi \Sigma$ channels show
invariant mass distributions peaking at this energy. We emphasize
that the reaction is a very clean one to produce this resonance,
free of contributions from $I=1$ sources.

We also take into account the $\eta_c p$ interaction, which is
enhanced close to a dynamically generated resonance $R$, from the
$\bar{D}\Sigma_c$, $\bar{D}\Lambda_c$, and $\eta_c N$ channels, due
to a relatively large coupling of the resonance to $\eta_c p$,
weaker than to $\bar{D}\Sigma_c$ (the largest component) but larger
than the coupling to the $\bar{D}\Lambda_c$ channel.

Up to a global normalization constant, we can compare the strength
of the reactions in the $K^ - p$ mass distribution close to the $K^-
p$ threshold, the strength of the $\pi^+ \Sigma^-$, $\pi^0
\Sigma^0$, and $\pi^- \Sigma^+$ mass distributions around the peak
of the upper $\Lambda(1405)$ state and the strength of the $\eta_c
p$ mass distribution at the peak of the $R$ resonance around $4265$
MeV. They all have a similar strength and should be easily
identifiable.

The results shown here are predictions for ongoing experiments at
LHCb, and comparison of the observed results with these predictions
will be most useful to pin down the different dynamical aspects of
hadron physics that we have discussed in this paper.

\section*{Acknowledgments}

We would like to thank L. Zhang for suggesting this problem. This
work is partly supported by the National Natural Science Foundation
of China (Grants No. 11475227, 11565007, 11647309, 11735003) and the
Youth Innovation Promotion Association CAS (No. 2016367). This work
is also partly supported by the Spanish Ministerio de Economia y
Competitividad and European FEDER funds under the contract number
FIS2011-28853-C02-01, FIS2011- 28853-C02-02, FIS2014-57026-REDT,
FIS2014-51948-C2- 1-P, and FIS2014-51948-C2-2-P, and the Generalitat
Valenciana in the program Prometeo II-2014/068. We acknowledge the
support of the European Community-Research Infrastructure
Integrating Activity Study of Strongly Interacting Matter (acronym
HadronPhysics3, Grant Agreement n. 283286) under the Seventh
Framework Programme of EU.

\section*{Appendix}

We write the operator responsible for the transition of Fig.~\ref{fig:fig1} and make the HQSS approach neglecting the terms of $1/m_{Q}$ ($m_{Q}$, the heavy quark mass). Considering the $W$ propagator as $D_W = g_{\mu \nu}/m^2_W$, we must evaluate matrix elements of the type
\begin{eqnarray}
t  = <c|\gamma^{\mu}(1-\gamma_5)|b><s|\gamma_{\mu}(1-\gamma_5)|c>.
\end{eqnarray}

Making the non relativistic reduction of the $\gamma_{\mu}$ and $\gamma_{\mu}\gamma_5$ matrices and keeping terms of order ${\mathcal O}(1)$ we must keep, $\gamma^0 \sim 1$ and $\gamma^i \gamma_5 \sim \sigma^i$ ($i =1$, 2, 3). Thus we have to evaluate the following matrix element
\begin{widetext}
\begin{eqnarray}
<S_1|1|M><M'|1|S_2>  -  \sum_{i=1}^3<S_1|\sigma^i|M><M'|\sigma^i|S_2>,  \label{eq:s1to2}
\end{eqnarray}
\end{widetext}
where $S_1$ and $S_2$ are the third components of the spins of the $c$, $\bar{c}$, and $M$ and $M'$ are the third spin components of the $b$ and $s$ quarks, respectively.

We next write
\begin{eqnarray}
\sigma^i \sigma^i \to \sum_{\mu}(-1)^{\mu}\sigma^{\mu}\sigma^{-\mu}
\end{eqnarray}
where $\sigma^{\mu}$ are the Pauli matrices in the spherical basis and using the Wigner Eckert theorem we have
\begin{widetext}
\begin{eqnarray}
<S_1|\sigma^\mu|M> = \mathcal{C}(\frac{1}{2} 1 \frac{1}{2}; M \mu S_1)<\frac{1}{2}||\sigma||\frac{1}{2}>
 = \sqrt{3} \mathcal{C}(\frac{1}{2} 1 \frac{1}{2}; M \mu S_1).
\end{eqnarray}
\end{widetext}

The other consideration is that we have to combine particle-antiparticle in angular momentum. We then take into account that the state $<J,-M|(-1)^{J+M}$ behaves like a state $|JM>$. Then we combine a state with $S_1$ and $-S_2$ to form $|jm>$, the $\eta_c$ or $J/\psi$ state with $j=0$ or $1$, respectively.

Then Eq.~\eqref{eq:s1to2} becomes
\begin{widetext}
\begin{eqnarray}
\sum_{S_1}(-1)^{\frac{1}{2}+S_1-m}\mathcal{C}(\frac{1}{2} \frac{1}{2} j; S_1,m-S_1) [ \delta_{S_1,M} \delta_{S_1-m, M'} - (-1)^{S_1 - M}3 \mathcal{C}(\frac{1}{2} 1 \frac{1}{2}; M,S_1-M)\mathcal{C}(\frac{1}{2} 1 \frac{1}{2}; S_1-m,-S_1+M,M') ] \nonumber
\end{eqnarray}
\end{widetext}
which implies in both terms $M' = M-m$, as it should be.

Next one reorders the Clebsch-Gordan coefficients to produce a Racah
coefficient~\cite{rose} as done in Ref.~\cite{Liang:2016ydj} and we
find
\begin{widetext}
\begin{eqnarray}
(-1)^{\frac{1}{2}+M-m} \mathcal{C}(\frac{1}{2} \frac{1}{2} j; M,m-M) [1-6W(\frac{1}{2}1j\frac{1}{2};\frac{1}{2}\frac{1}{2})]
= (-1)^{\frac{1}{2}+M-m} \mathcal{C}(\frac{1}{2} \frac{1}{2} j; M,m-M) C,   \label{eq:CGC}
\end{eqnarray}
\end{widetext}
with $C=-2$ and $2$ for $j=0$ and $1$, respectively.

In addition one has the radial matrix element
\begin{eqnarray}
\frac{1}{4\pi}\int r^2 dr \phi_b(r) \phi_{c1}(r) \phi_{c2}(r) \phi_s(r) j_0(qr),
\end{eqnarray}
with $c_1$ and $c_2$ corresponding to the $c$, $\bar{c}$ quarks and $b$, $s$ to the $b$, $s$ quarks, while $q$ is the momentum transfer and we have assumed that the $s$ quark is in $l=0$, as if we were producing the $\Lambda$ ground state.

When we produce $\eta_c K^- p$ with $K^- p$ in $S$-wave, the final state $K^- p$ must be obtained from the hadronization, as shown in Fig.~\ref{fig:fig2}, but since $K^- p$ in $S$-wave has negative parity the $s$ quark prior to the hadronization must have negative parity because the $ud$ quark pair is spectator and has positive parity. Then one has to have the $s$ quark excited to $l$ odd and we take $l =1$, the lowest one, which leads to $J=1/2$ that one has with $K^- p$ in $S$-wave. Eq.~\eqref{eq:CGC} is generalized in this case and we find
\begin{widetext}
\begin{eqnarray}
(-1)^{\frac{1}{2}+M-m} \mathcal{C}(1 \frac{1}{2} J;M'- M+m,M-m,M') \mathcal{C}(\frac{1}{2} \frac{1}{2} j; M,m-M) C, \label{eq:mm}
\end{eqnarray}
\end{widetext}
where $J$ is the $s$ total spin that comes from the combination of spin and the $l$ angular momentum of the $s$ quark, and $M'$ its third component. The radial matrix element now becomes
\begin{eqnarray}
\frac{1}{\sqrt{4\pi}} (-1)^l Y^*_{l0}(\hat q)\int r^2 dr \phi_b(r) \phi_{c1}(r) \phi_{c2}(r) \phi_s(r) j_l(qr). \nonumber
\end{eqnarray}

The next step is to combine $|jm>$ with $|1/2,M-m>$ to give the
initial $|\frac{1}{2},M>$ state in the case of $s$ quark with $l=0$,
multiplying by the Clebsch-Gordan coefficient ${\mathcal
C}(j\frac{1}{2}\frac{1}{2};m,M-m)$ and summing over $m$. The
resulting amplitude becomes
\begin{eqnarray}
(-1)^{1+j}\big( \frac{2j+1}{2} \big )^{1/2}C,
\end{eqnarray}
which indicates that the $J/\psi$ production in this case would be
three times bigger than for $\eta_c$.

On the other hand, in the case we are concerned about, with the $s$
quark in $l=1$, we must combine $|jm>$ with $|JM'>$ to give
$|\frac{1}{2}M>$, multiplying by the Clebsch-Gordan coefficient
${\mathcal C}(jJ\frac{1}{2};m,M',M)$ and summing over $m$. This
makes $M' = M-m$. Once again we recombine the three Clebsch-Gordan
coefficients into one Clebsch-Gordan and one Racah coefficient,
$W(1\frac{1}{2}\frac{1}{2}j;\frac{1}{2}\frac{1}{2})$, with the final
result for the amplitude for $J=1/2$ ($K^- p$ in $S$-wave),
\begin{eqnarray}
\mathcal{C}(1\frac{1}{2}\frac{1}{2};0M)C',
\end{eqnarray}
with $C' = \sqrt{2}$ for $j=0$ and $C' = - \frac{\sqrt{6}}{3}$ for
$j=1$. Since $|\mathcal{C}(1\frac{1}{2}\frac{1}{2};0M)|^2$  is
independent of $M$, the probability to production $\eta_c K^-p$ in
$S$-wave is now three times bigger than for $J/\psi K^-p$.

\bibliographystyle{plain}

\end{document}